\documentclass[aps,prc,noshowpacs,superscriptaddress,amsmath,amssymb,
  twocolumn,preprintnumbers]{revtex4}
\usepackage{amsmath,amssymb,lmodern}
\usepackage{bbold}
\usepackage{bm}
\usepackage{relsize}
\usepackage{palatino}
\usepackage{braket}
\usepackage{bigints}
\usepackage{graphicx}
\graphicspath{{figures/}}
\usepackage{xcolor}

\newcommand{\tit}{\textit}

\newcommand{\tat}{\textit}

\DeclareMathOperator{\Tr}{Tr}

\newcommand{\average}[1]{\langle{#1}\rangle}

\begin{document}
\title{Thermal and hard scales in transverse momentum distributions,
  fluctuations and entanglement}
\author{X. Feal}\email{xgarciafe@bnl.gov}
\affiliation{Physics Department, Brookhaven National Laboratory, Bldg. 510A, Upton, NY 11973, USA}
\affiliation{ IGFAE/Instituto Galego de F\'{\i}sica de Altas
  Enerx\'{\i}as \& \\ Universidade de Santiago de Compostela, 15782
  Santiago, SPAIN }
\author{C. Pajares}\email{pajares@igfae.usc.es}
\author{R.A. Vazquez}\email{vazquez@igfae.usc.es}
\affiliation{ IGFAE/Instituto Galego de F\'{\i}sica de Altas
  Enerx\'{\i}as \& \\ Universidade de Santiago de Compostela, 15782
  Santiago, SPAIN }

\date{\today}
\begin{abstract}
  We analyze the transverse momentum distributions of \textit{pp, pPb, XeXe}
  and \textit{PbPb} collisions at different RHIC and LHC energies and
  centralities as well as the corresponding distributions for Higgs production
  decaying into $\gamma\gamma$ and 4\textit{l}. A simple linear relation
  is found between the effective thermal temperature and the hard scale,
  approximately valid for all processes and mainly determined by the hard
  scale fluctuations. In order to go further, it is shown that the whole
  spectrum of \textit{pp} collisions can be described by a single function
  showing that the thermal temperature is determined solely by the hard scale
  and its fluctuations. The possible relation between the multiplicities of
  the soft and hard scales is explored. 
\end{abstract}

\maketitle

\section{Introduction}

The apparent thermal features of proton-proton collisions is a challenge to
understand the collective behavior observed in small systems, where the
application of the conventional hydrodynamical explanation is questionable
\cite{yi2019,strickland2019}. On the other hand, theoretical studies of
quenches in entangled systems described by (1+1)-dimensional conformal field
theories of expanding quantum fields and strings, show that the system obeys a
generalized Gibbs ensemble with an effective temperature set by the energy
cut-off for the ultraviolet modes
\cite{calabrese2005,calabrese2016,bergers2018a,bergers2018b}.  In the last
years there has been a large activity in the field
\cite{Tu,Kovner,Armesto,Gotsman_1,Gotsman_2,Gotsman_3,Giannini,Iskander,Castorina,Afik,Ramos}.

In a high energy collision a hard parton interaction produces a rapid quench
of the entangled partonic state \cite{kharzeev2017} and thus the corresponding
effective temperature, inferred from the exponential shape of the transverse
momentum distribution (TMD) of the secondaries of the collision, can depend on
the scale of the collision, which works as an ultraviolet cut-off of the
quantum modes resolved by the collision. This possibility was recently studied
in proton-proton collisions at different energies and multiplicities, in the
production of the Higgs boson and in \textit{PbPb} collisions, showing that in
fact there is a relation between the hard scale and the effective temperature
\cite{baker2018,feal2019,bellwied2018}.

In this paper we perform an extensive study of the energy and multiplicity
dependence in \textit{pp} collisions
\cite{rhic2009,rhic2006,alice2013,alice2018a,alice2018b,alice2016} as well as
\textit{pPb} \cite{alice2016b}, \textit{XeXe} \cite{alice2019} and
\textit{PbPb} \cite{alice2018a} collisions, showing that the relation between
both scales is determined approximately by the inverse of the normalized
fluctuations of the number of partons of the initial wave function or,
equivalently, of the normalized fluctuations of the hard scale. The hard
process, with transverse momentum $p_\perp$, probes only the region of the
space $H$ of transverse size $1/p_\perp$. Let us denote by $S$ the region of
space complementary to $H$. The initial state is described by the wave
function
\begin{align}
  |\Psi\rangle = \sum_n \alpha_n |\Psi_n^H\rangle\otimes| \Psi_n^S\rangle,
  \label{entangled_state}
\end{align}
of a suitably chosen orthonormal set of states $|\Psi_n^H\rangle$ and $|
\Psi_n^s\rangle$ localized in the domains $H$ and $S$, with different numbers
$n$ of partons. The state \eqref{entangled_state} cannot be separated into a
product $|\Psi^H\rangle\otimes|\Psi^S\rangle$, and therefore
$|\Psi\rangle$ is entangled. The density matrix of the mixed state probed
in the region $H$ is
\begin{align}
  \rho_H &= \Tr_S \rho = \sum_n \langle \Psi_n^S|\Psi\rangle \langle
  \Psi|\Psi_n^S\rangle = \sum_n |\alpha_n|^2|\Psi_n^H\rangle \langle \Psi_n^H|,
\end{align}
where $|\alpha_n|^2\equiv p_n$ is the probability of having a state with $n$
partons. We can consider that a high momentum partonic configuration of the
initial state when the interaction takes place undergoes a rapid quench. The
onset $\tau$ of this hard interaction is given by the hardness scale,
$\tau\sim 1/p_\perp$. Because $\tau$ is small the quench creates a highly
excited multi-particle state. The produced particles have a thermal like
exponential spectrum with an effective temperature which is determined by the
hard scale and the fluctuations on the number of partons.
With these considerations, we fit the different TMDs by an exponential
distribution and a power like distribution
\cite{bylinkin2014,baker2018,feal2019},
\begin{align}
\frac{1}{N_{ev}}\frac{1}{2\pi p_\perp}\frac{d^2N_{ev}}{d\eta
  dp_\perp}=A_{th}e^{-m_\perp/T_{th}} +\frac{A_h}{\big(1+m_\perp^2/kT_h^2\big)^k},
\label{pheno_fit}
\end{align}
where $T_{th}$ is the effective temperature, $T_h$ is the hard scale and $k$ a
parameter which is determined by the falloff of the different distributions at
high $p_\perp$.
\begin{table}
\caption{Temperatures and falloff $k$ of the TMD of charged particles in
  \textit{pp} collisions at different RHIC and LHC colliding energies.}
 \label{tab:table_1}
  \begin{tabular}{@{} c c c c c c@{}}
    \\\hline\hline 
& $\sqrt{s_{nn}}$ & T$_{th}$(GeV)  & T$_{h}$(GeV) & $k$ &\\
\hline
& 64 GeV & 0.188 & 0.719 & 5.50 &\\
& 200 GeV & 0.189 & 0.852 & 4.62 & \\
& 900 GeV & 0.176 & 0.703 & 3.62 & \\
& 2.76 TeV & 0.180 & 0.713 & 3.28 &\\
& 5.02 TeV & 0.184 & 0.735 & 3.13 &\\
& 7 TeV & 0.180 & 0.716 & 3.08 &\\
& 13 TeV & 0.181 & 0.744 & 3.01 &\\
\hline\hline
  \end{tabular}
\end{table}
\begin{figure}
\includegraphics[scale=0.65]{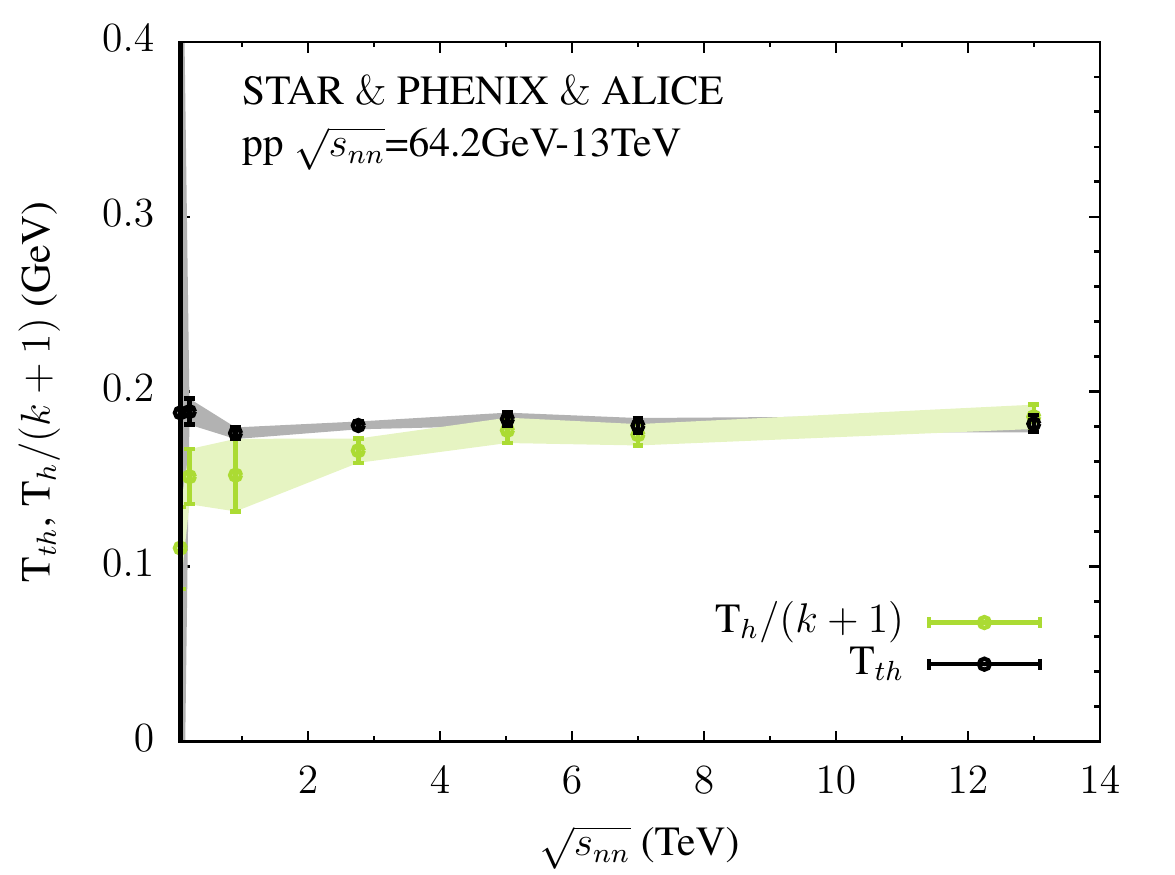}
\caption{(color online) Thermal scale T$_{th}$ compared to the quenched hard
  scale T$_{h}/(k+1)$, extracted from RHIC and LHC collected data on \tit{pp}
  collisions, as a function of the collision energy.}
\label{fig:figure_1}
\end{figure}

In the next sections, we will show that the relation between the two
scales is given by the size of the fluctuations of the hard
scale. This relation is universal, valid for \tit{pp}, \tit{pPb},
\tit{XeXe}, and \tit{PbPb} for all energies and centralities. In the
case of \tit{pp} collisions, we are able to describe the whole
$p_\perp$ spectrum with a single function for energies ranging from
RHIC to LHC. This function provides us with the low $p_\perp$
fluctuations needed to improve the universal agreement found in
\tit{AA} and \tit{pA} collisions.

\section{The thermal temperature determined by the hard scale and its
  fluctuations}
\label{sect:thermal}

The results of our fit of the TMDs to \eqref{pheno_fit} are shown in
tables, and the corresponding uncertainties of the extracted
parameters are shown as error bands in the accompanying figures. In
Table \ref{tab:table_1}, we show the values of $T_{th}$, $T_h$ and $k$
obtained from the fits to the TMD of charged particles produced in
$pp$ collisions at very different RHIC and LHC energies.  In Table
\ref{tab:table_2}, we show the results from the fits to the TMD of
charged pions for several multiplicity bins in $pp$ collisions at
$\sqrt{s_{nn}}$=7 TeV, in the range $|\eta|<0.5$ and
0.1$<p_\perp<$17.5 GeV/c. The obtained values of temperature are
larger for charged particles than for pions, as expected.
In Table \ref{tab:table_3}, we show the fit results for \tit{pPb} collisions at
$\sqrt{s_{nn}}$=5.02 TeV in the range -0.5$<\eta<$0 and 0.1$<p_\perp<$17.5
GeV/c. In Table \ref{tab:table_4}, we show the fit results for \tit{XeXe}
collisions at $\sqrt{s_{nn}}$=5.44 TeV at different centralities in the range
$\eta<$0.8 and 0.2$<p_\perp<$17 GeV/c, and in Table \ref{tab:table_5} for
\tit{PbPb} collisions at $\sqrt{s_{nn}}$=5.02 TeV at different centralities in
the range $\eta<$0.8 and 0.2$<p_\perp<$17 GeV/c.
\begin{table}
  \caption{Characteristic temperatures and falloff $k$ of the TMD of
    $\pi^{\pm}$ in \textit{pp} collisions at $\sqrt{s_{nn}}$=7 TeV for several
    multiplicities.}
 \label{tab:table_2}
  \begin{tabular}{@{} c c c c c c@{}}
    \\\hline\hline 
    & $dN_{c}/d\eta$ & T$_{th}$(GeV)  & T$_{h}$(GeV) & $k$ &\\
\hline
& 21.3 & 0.147 & 0.675 & 2.99 &\\
& 16.5 & 0.146 & 0.660 & 3.00 & \\
& 13.5 & 0.146 & 0.652 & 3.03 & \\
& 11.5 & 0.146 & 0.637 & 3.03 &\\
& 10.1 & 0.146 & 0.633 & 3.04 &\\
& 8.45 & 0.146 & 0.620 & 3.05 &\\
& 6.72 & 0.146 & 0.608 & 3.06 &\\
& 5.4 & 0.148 & 0.598 & 3.08 &\\
& 3.9 & 0.150 & 0.589 & 3.14 &\\
& 2.3 & 0.154 & 0.553 & 3.23 &\\
\hline\hline
  \end{tabular}
\end{table}
\begin{figure}
\includegraphics[scale=0.65]{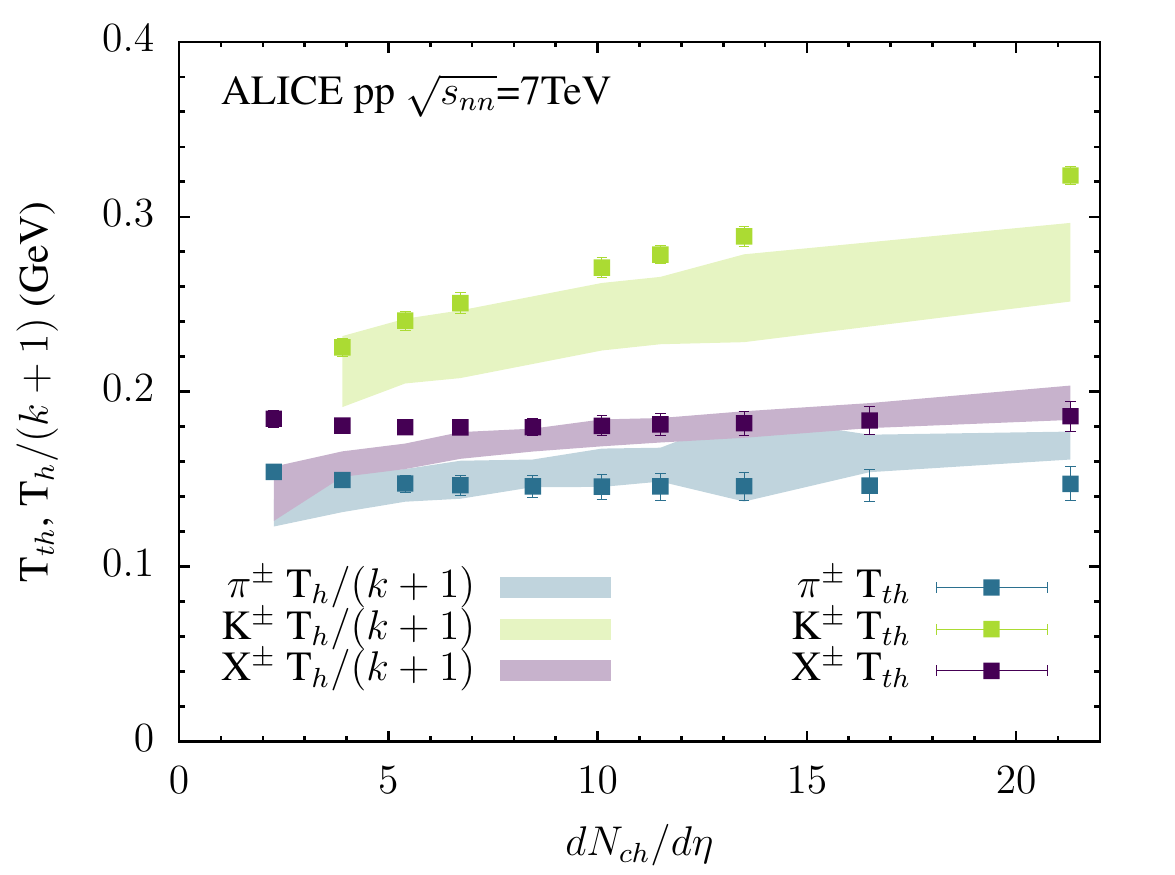}
\caption{(color online) Thermal scale T$_{th}$ compared to the quenched hard
  scale T$_{h}/(k+1)$, extracted ALICE collected data on \tit{pp} collisions at
  $\sqrt{s_{nn}}$=7 TeV, as a function of the charged particle multiplicity.}
\label{fig:figure_2}
\end{figure}
\begin{table}
  \caption{Characteristic temperatures and falloff $k$ of the TMD of
    $\pi^{\pm}$ in \textit{pPb} collisions at $\sqrt{s_{nn}}$=5.02 TeV for
    several multiplicities.}
 \label{tab:table_3}
  \begin{tabular}{@{} c c c c c c@{}}
    \\\hline\hline 
    & $dN_{ch}/d\eta$ & T$_{th}$(GeV)  & T$_{h}$(GeV) & $k$ &\\
\hline
& 45.0 & 0.156 & 0.709 & 3.23 &\\
& 36.2 & 0.154 & 0.718 & 3.23 & \\
& 30.5 & 0.152 & 0.705 & 3.18 & \\
& 23.2 & 0.150 & 0.699 & 3.18 &\\
& 16.1 & 0.148 & 0.678 & 3.15 &\\
& 9.8 & 0.148 & 0.655 & 3.12 &\\
& 4.3 & 0.150 & 0.613 & 3.11 &\\
\hline\hline
  \end{tabular}
\end{table}
\begin{figure}[ht]
\includegraphics[scale=0.65]{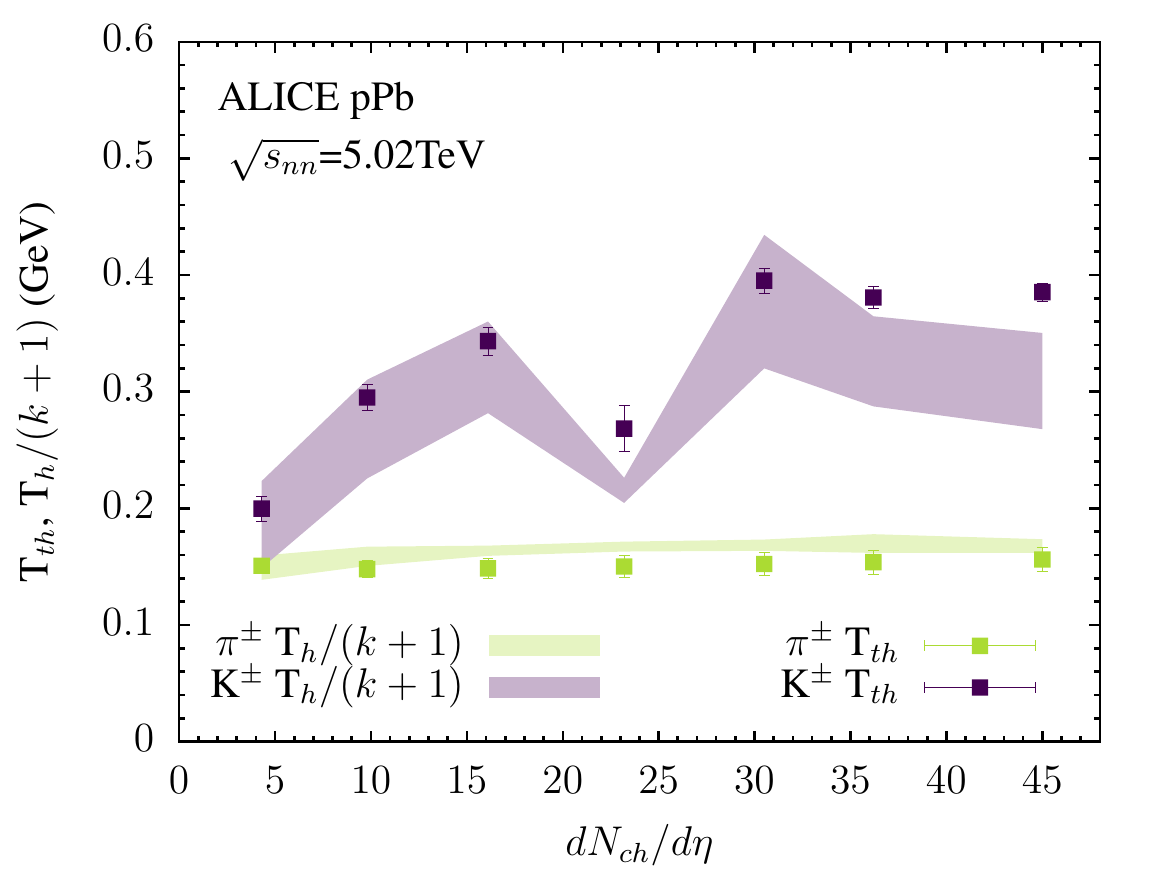}
\caption{(color online) Thermal scale T$_{th}$ compared to the quenched hard
  scale T$_{h}/(k+1)$, extracted from ALICE collected data on \tit{pPb}
  collisions at $\sqrt{s_{nn}}$=5.02 TeV, as a function of the charged
  particle multiplicity.}
\label{fig:figure_3}
\end{figure}
\begin{table}
  \caption{Characteristic temperatures and falloff $k$ of the TMD of charged
    particles in \textit{XeXe} collisions at $\sqrt{s_{nn}}$=5.44 TeV for
    several centrality classes.}
 \label{tab:table_4}
  \begin{tabular}{@{} c c c c c c@{}}
    \\\hline\hline 
    & $dN_{ch}/d\eta$ & T$_{th}$(GeV)  & T$_{h}$(GeV) & $k$ &\\
\hline
& 1167 & 0.138 & 0.617 & 3.36 &\\
& 939 & 0.136 & 0.637 & 3.36 & \\
& 706 & 0.135 & 0.636 & 3.33 & \\
& 478 & 0.132 & 0.620 & 3.27 &\\
& 315 & 0.129 & 0.612 & 3.21 &\\
& 198 & 0.126 & 0.611 & 3.19 &\\
& 118 & 0.123 & 0.619 & 3.18 &\\
& 65 & 0.119 & 0.602 & 3.13 &\\
& 32 & 0.114 & 0.615 & 3.17 &\\
\hline\hline
  \end{tabular}
\end{table}
\begin{table}
  \caption{Characteristic temperatures and falloff $k$ of the TMD of charged
    particles in \textit{PbPb} collisions at $\sqrt{s_{nn}}$=5.02 TeV for
    several centrality classes.}
 \label{tab:table_5}
  \begin{tabular}{@{} c c c c c c@{}}
    \\\hline\hline 
    & $dN_{ch}/d\eta$ & T$_{th}$(GeV)  & T$_{h}$(GeV) & $k$ &\\
\hline
& 1942 & 0.140 & 0.598 & 3.41 &\\
& 1585 & 0.139 & 0.606 & 3.39 &\\
& 1180 & 0.138 & 0.612 & 3.36 &\\
& 786 & 0.135 & 0.613 & 3.31 &\\
& 512 & 0.132 & 0.614 & 3.27 &\\
& 318 & 0.129 & 0.606 & 3.22 &\\
& 183 & 0.125 & 0.606 & 3.19 &\\
& 96 & 0.120 & 0.602 & 3.17 &\\
& 45 & 0.116 & 0.574 & 3.10 &\\
\hline\hline
  \end{tabular}
\end{table}
\begin{figure}[ht]
\includegraphics[scale=0.65]{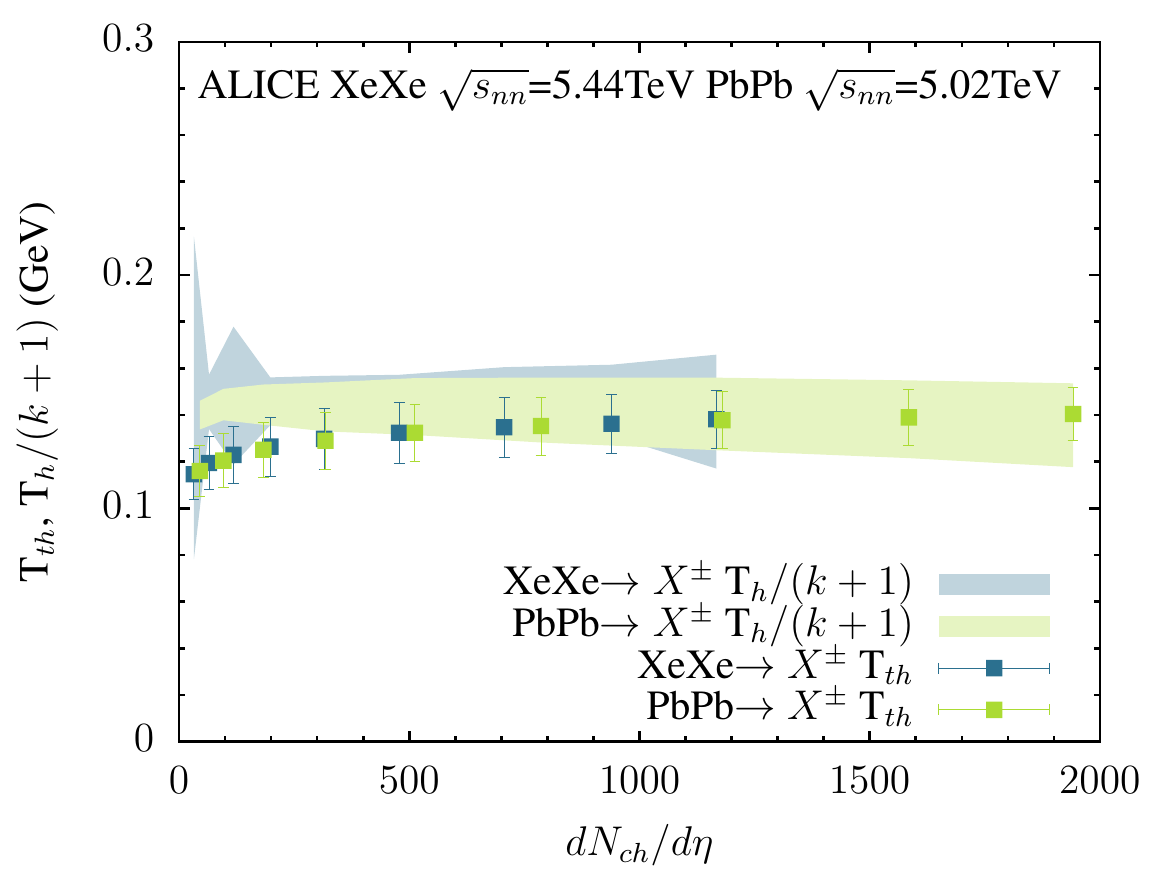}
\caption{(color online) Thermal scale T$_{th}$ compared to the quenched hard
  scale T$_{h}/(k+1)$, extracted from ALICE collected data on \tit{XeXe}
  collisions at $\sqrt{s_{nn}}$=5.44 TeV and \tit{PbPb} collisions at
  $\sqrt{s_{nn}}$=5.02 TeV, as a function of the charged particle
  multiplicity.}
\label{fig:figure_4}
\end{figure}
The general behavior of the temperatures $T_{th}$ and $T_h$ is to increase
with the colliding energy, as expected. In the case of the hard scale $T_h$,
it also increases with the centrality in all cases. However, $T_{th}$ smoothly
decreases with the centrality in \tit{pp} collisions, contrary to the rest of
cases. The behavior of the falloff index $k$ with centrality is the same as
the thermal temperature $T_{th}$, it increases with centrality except in
\tit{pp} collisions, where it decreases.
%
%
From these values we have found the approximate relation
\begin{align}
  \frac{T_h}{k+1}=T_{th}.
  \label{thardtthermalrelation}
\end{align}
In Figure \ref{fig:figure_1} we plot $T_{th}$ and $T_h/(k+1)$ for \tit{pp}
collisions at different energies. In Figures \ref{fig:figure_2},
\ref{fig:figure_3}, and  \ref{fig:figure_4} we plot the cases
of \tit{pp}, \tit{pPb}, \tit{XeXe}, and \tit{PbPb} collisions at
$\sqrt{s_{nn}}$=7 TeV, 5.02 TeV, and  5.44 TeV
for different identified charged particles as a function of the
multiplicity. We have looked at the transverse momentum distribution of Higgs
production decaying into $\gamma\gamma$ and 4$l$, as reported in reference
\cite{baker2018}, obtaining $T_{th}$=3.5$\pm$0.7 GeV, $T_h$=14.4$\pm$0.3 and
$k$=3.7$\pm$0.4. Hence we obtain a quenched hard scale $T_h/(k+1)$=3.1$\pm$0.4
GeV, to compare with $T_{th}$=3.5$\pm$0.7 GeV. In the same line, studies of
the $p_\perp$ distribution of the W's produced in $p\bar{p}$ collisions have
found similar behaviors \cite{albajar1989,fletcher1990}.

An overall agreement is observed in all cases, except in very low
multiplicity events and very low colliding energies in \tit{pp}
collisions, where the discrepancies between both quantities becomes
substantial. This agreement is remarkable, considering the large
number of TMDs studied and the large differences presented in the
temperatures for different projectiles and targets, as well as
centralities and energies.

\begin{figure}
\includegraphics[scale=0.65]{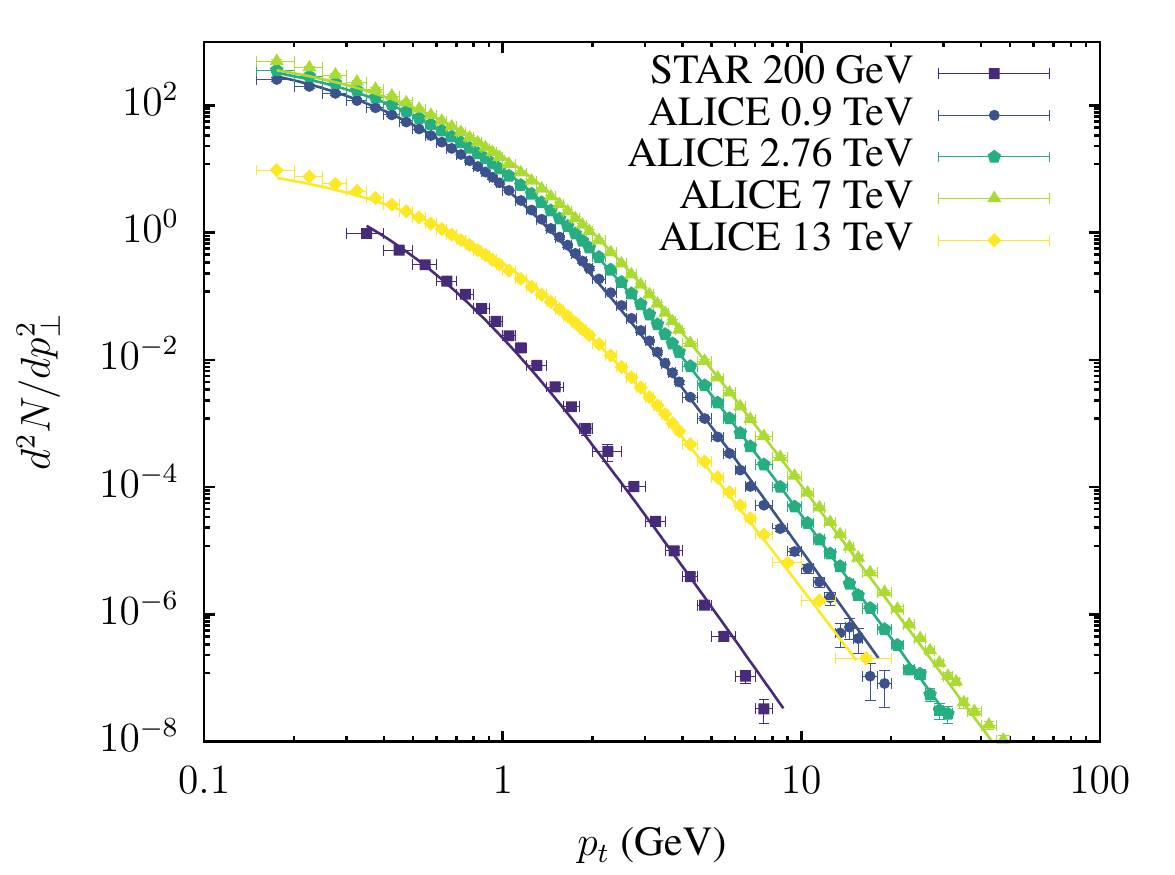}
\caption{(color online) $d^2N/dp_\perp^2$ as a function of  $p_\perp$
  for \textit{pp} collisions at
  different energies (symbols, as marked) together with the fit to the
  hypergeometric function (lines). Data for ALICE at 13 TeV is for
  $d^2\sigma/dp_\perp^2$}
\label{fig:figure_5}
\end{figure}

In order to improve the agreement with $pp$ data, we devise now a
simple expression describing well the whole soft and hard spectrum for
the full range of energies explored at RHIC and the LHC. The hard
part of the TMD \eqref{pheno_fit} can be rewritten as
\begin{align}
  \frac{1}{\big(1+p_\perp^2/\gamma\big)^k}&=\int^\infty_0 dx
  e^{-p_\perp^2x}\frac{\gamma}{\Gamma(k)}(\gamma x)^{k-1}
  e^{-\gamma x}\nonumber\\
  &=\int_0^\infty dx f(p_\perp,x)W_p(x),\label{gamma_convolution_tmd}
\end{align}
where $W_p(x)$ is the Gamma distribution and thus $1/k$ can be
understood as the normalized fluctuations of the hard scale $T_h$
\begin{align}
\frac{1}{k} = \frac{\langle x^2 \rangle -\langle x \rangle^2}{\langle x
  \rangle^2}, \medspace\medspace\medspace\medspace\medspace\medspace
\medspace\medspace
\gamma\equiv kT_h^2 .
\end{align}
We can add an additional source of fluctuations for $T_h$ using a
Gaussian distribution $G(T_h)$. The whole $p_\perp$ distribution is
given by
\begin{align}
  F(p_\perp) = & \int_0^\infty dT_h G(T_{h}) \frac{1}{(1+p_\perp^2/\gamma)^k}
  \nonumber
  \\
   =  & 
  \int_0^\infty dT_h G(T_{h}) \int_0^\infty dx W_p(x) e^{-p_\perp^2x},
\end{align}
The above equation can be now cast into
\begin{equation}
  F(p_\perp) = \int_0^\infty dx W(x) e^{-p_\perp^2 x},
  \label{newpt}
\end{equation}
where now $W(x)$ can be approximated by
\begin{equation}
  W(x) = \mathcal{N} \frac{x^{k-1}}{(1+x/\bar x)^{k'}},
\end{equation}
where $\mathcal{N}$ is a normalization constant $\bar x $ is the scale
of $x$ and $k'= k+1/2$ is required to obtain Gaussian fluctuations at
high $x$. The above equation has the correct asymptotic behavior for
both low and high $x$.  In this way the transverse momentum
distribution \eqref{newpt} becomes
\begin{equation}
  F(p_\perp) = \mathcal{N} \bar x^k \; \Gamma(k) \; U(k;1/2;\bar x p_\perp^2).
  \label{newpt_exp}
\end{equation}
$U$ is the confluent hypergeometric function. Its asymptotic
limits are for $p_\perp\rightarrow 0$
\begin{equation}
  F(p_\perp) = C \left(1 - \frac{2 \Gamma(k+1/2)}{\Gamma(k)} \bar x^{1/2} \,
  p_\perp  \right),
\end{equation}
and for $p_\perp\rightarrow \infty$
\begin{equation}
  F(p_\perp) = \frac{C'}{(\bar x p_\perp^2)^k}.
\end{equation}
At high $p_\perp$, a power like behavior is obtained with power $2k$, and at
low $p_\perp$ the thermal behavior $\exp(-p_\perp/T_{\rm th})$ with
\begin{equation}
T_{\rm th} = \frac{\Gamma(k)}{2\Gamma(k+1/2)} \frac{1}{\bar x^{1/2}}.
\end{equation}
As the fluctuations encoded in the function $W(x)$ depend only on the
scale $\bar x$ and $k$, we can say that the effective thermal
temperature depends only on the scale $\bar x$ and its fluctuations.
In Figure \ref{fig:figure_5}, we show the fit using the whole soft and
hard spectrum \eqref{newpt_exp} to \textit{pp} data at different
energies. A good description is obtained in all cases.

\section{Multiplicity distribution scales}
\label{sect:CSPM}

In what concerns to the multiplicity distribution, we note that a gamma
distribution on the number of partons is also obtained for events which have
at least one high $p_\perp$ particle due to a hard parton collision. In fact,
if $P(n)$ is the probability of having $n$ partons in a given collision, the
probability $P^c(n)$ of having $n$ partons with at least one hard is
\cite{diasdedeus1997a,diasdedeus1997b}
\begin{align}
  P^c(n)=\frac{n}{\langle n \rangle} P(n).
\end{align}
This selection procedure can be repeatedly applied forming the chain
\begin{align}
  P(n)\to \frac{n}{\langle n \rangle} P(n) \to \frac{n^2}{\langle n^2\rangle}
  P(n) \to \frac{n^k}{\langle n^k\rangle} P(n).
\end{align}
%
%
Similarly, we also notice that a gamma distributed multiplicity
density convoluted with a Poisson process, produces a negative
binomial distribution (NBD) for the multiplicity, broadly used to
describe the experimental data. Namely
\begin{align}
  \frac{\Gamma(n+k_n)}{\Gamma(n+1)\Gamma(k_n)}&
  \frac{\gamma_n^{k_n}}{(1+\gamma_n)^{k_n+n}}
  =\int^\infty_0 dN
  \frac{e^{-N}N^n}{n!}\label{gamma_convolution_multiplicity}\\
  \times(\gamma_nN)^{k_n-1}&\exp(-\gamma_nN)=\int^\infty_0 dN P(n,N)W_n(N),
  \nonumber
\end{align}
where as before
\begin{align}
\frac{1}{k_n}=\frac{\langle N^2\rangle-\langle N\rangle^2}{\langle N
  \rangle^2},
\medspace\medspace\medspace\medspace\medspace\medspace\medspace\medspace
\gamma_n\equiv\frac{k_n}{\langle N \rangle}.
\end{align}
Since the mean multiplicity and the $p_\perp$ distribution are related as follows
\begin{align}
\langle n \rangle = \int d^2p_\perp \int dx f(p_\perp,x) W_p(x),
\end{align}
the following relation between the two gamma distributions can be written
\begin{align}
  W_p(x)=\frac{\gamma_n}{\gamma}W_n(x),
  \label{gamma_relation}
\end{align}
with $k=k_n+2$. The convolution of the gamma distribution with a
Poisson distribution gives rise to a negative binomial distribution
for the multiplicity distribution, see formula
\eqref{gamma_convolution_multiplicity}. As far as the gamma
distribution is obtained for at least one hard parton, the resulting
multiplicity distribution describes the multiplicity distribution
events with at least one hard parton. For the rest of events another
distribution is required. There are several fits to the \tit{pp} data
at different colliding energies using two negative binomial
distributions
\cite{alkin2017,alice2017,zborovsky2018,biyajima2019}. Each of these
distributions has two parameters, $k$ fixing the fluctuations and
$\langle n \rangle$ the mean multiplicity. The two mean multiplicities
can be seen as the two multiplicity scales corresponding to the two
transverse momentum scales $T_{th}$ and $T_h$, and we can look for a
relation between these scales similar to the relation
\eqref{thardtthermalrelation}.

According to formula \eqref{gamma_relation}, the parameter $k$ of the gamma
distribution is two units larger than the one corresponding to the gamma
distributions on the number of partons, thus we expect that instead of the
equation \eqref{thardtthermalrelation} for the $p_\perp$ distribution, we
should have
\begin{align}
\frac{\langle N_h \rangle}{k-1} = \langle N_s\rangle.
\end{align}
This equation must be seen with caution because in the case of
multiplicities longitudinal momentum fluctuations add to the $p_\perp$
fluctuations. In order to avoid these contributions we look at the
data on small rapidity range. In Table \ref{tab:table_6} we show the
results of a fit \cite{alkin2017} using two negative binomial
distributions to the multiplicity distributions of \textit{pp}
collisions in the pseudo-rapidity range $\left| \eta \right|\le$ 0.5
for different energies. The comparison of columns two and four shows a
reasonable agreement.

\begin{table}
  \caption{Relation between the mean multiplicities for events having at least
    one hard parton and the mean multiplicity for the rest of the events
    extracted from \cite{alkin2017}.}
 \label{tab:table_6}
  \begin{tabular}{@{} c c c c c c@{}}
    \\\hline\hline 
    & $\sqrt{s_{nn}}$(TeV) & $\langle N_s\rangle$  &
    $\langle N_h\rangle$ & $\langle N_h\rangle/(k-1)$\\
\hline
& 0.9 & 2.1$\pm$1.9 & 5$\pm$4 & 1.9$\pm$1.5 &\\
& 2.76 & 2.5$\pm$1.0 & 7$\pm$2 & 3.1$\pm$0.9 &\\
& 7 & 3.6$\pm$1.4 & 12$\pm$3 & 5.8$\pm$1.4 &\\
\hline\hline
  \end{tabular}
\end{table}

\section{Discussion}

In the left hand side of equation \eqref{thardtthermalrelation}, $T_h$
and $k$ are parameters related to hard collisions and thus described
by perturbative QCD. On the other hand $T_{th}$ has to do with
non-perturbative QCD. Thus, to some extent, equation
\eqref{thardtthermalrelation} links perturbative and non-perturbative
physics. The factor $k+1$ in equation
\eqref{thardtthermalrelation} determines the fall-off of the
probability of having an additional hard parton normalized by the hard
scale $T_h^2$. Such probability is just the variation of the hard
transverse momentum distribution.

Relations between these two regimes have been recently put forward in
different quantum problems \cite{QuantumProblems} and, more recently,
have been suggested as entropy constraints in an entangled nucleon,
relating the final state multiplicity of the fragmenting nucleon with
the parton distribution function probed by hard processes in $pp$ and
$ep$ collisions \cite{Tu}. Testing the implications of this
entanglement, the H1 collaboration has measured very recently the
charged particle distribution in DIS at HERA \cite{H1:2020zpd}. The
hadron entropy found in data does not confirm, however, these
entanglement predictions.

It is convenient to discuss if this disagreement is related to the way
in which the entanglement entropy of the nucleon has to be
obtained. There is not, to our knowledge, a known way of computing
from first principles the distribution of weights in the entangled
nucleon. Cascade models not including saturation and non-linear
evolution \cite{kharzeev2017} may not be sensitive enough to correctly
describe the initial entanglement entropy. Thus the observed entropy
in the multiplicity distribution of hadrons may not be a in an
one-to-one correspondence with the entropy of the cascade. 

However, we notice that the observed multiplicity distributions in
data are well described by NBDs. We can devise then a way of
reconstructing the cascade process under these phenomenological
considerations. We may assume the weights of the entangled state to be
formed according to a Poisson process, with a given mean value
$\lambda$. Then each of these partons gives rise to a cascade with a
geometric distribution \cite{kharzeev2017}, with parameter $p =
\average{n}/(\average{n}+k_n)$ ($k_n=1$ for a geometric distribution).
For the case of DIS at the energies explored
by the H1 collaboration \cite{H1:2020zpd}, the fall-off parameter
$k_n$ is large enough to transform back the final NBD to the initial
Poisson distributed multiplicity. Following this observation, $p$ has
to be small and the final state multiplicity mirrors the
initial weight distribution instead of being geometrically
distributed. At larger energies, however, $p$ must increase so that
the part of the cascade becomes more important.

At the level of the Von Neumann entropy, this energy evolution
corresponds to the passage from a Poisson distributed entropy to a
geometric/Gamma distributed entropy 
\begin{align}
   \log \langle n\rangle^{1/2} \to \log \langle n\rangle
\end{align}
the fall-off $k$ of the NBD interpolating between these two limits
\cite{feal2019}. At LHC energies for \textit{pp} collisions we expect
that the parameter $p$ is large enough that the entropy is dominated by the
geometric term and $S \propto \log  \langle n\rangle$, as was pointed out in
\cite{Tu}. At even larger energies
due to saturation of partons we expect to recover a behavior $S \propto \log
\langle n\rangle^{1/2}$. 
Accordingly, the number of micro-states is not anymore
$n$, but saturates as $\sqrt{n}$ when the colliding energy increases,
following the expectations of the glasma picture of the CGC
\cite{McLerran:1993ni} or the string percolation model
\cite{DiasdeDeus2011}.
 
\section{Conclusions}

Summarizing up, the analysis of the transverse momentum distributions
of \tit{pp}, \tit{pPb}, \tit{XeXe} and \tit{PbPb} collisions at
different RHIC and LHC energies and centralities together with Higgs
production decaying into $\gamma\gamma$ and 4l suggest that a hard
collision provides an ultraviolet scale that quenches the spectrum by
means of fluctuations of the hard scale. A simple relation between the
effective temperature and the hard scales is obtained which is
approximately satisfied in the different cases in study. In this way,
a non-perturbative scale $T_{th}$ has been related to two perturbative
quantities, $T_h$ and $k$. A gamma distribution is found, in agreement
with phenomenological descriptions, for the distribution of the hard
scale as well as the number of partons. The normalized fluctuations of
both distributions are related and give rise to a relation between the
multiplicities of the soft and hard spectrum. These findings are in
line with the possibility that a hard parton collision works as an
ultraviolet cutoff producing a quench of the rest of the entanglement
partons of the initial wave function. Such entanglement may be at the
origin of the apparent thermalization of the colliding hadrons.

\section{Acknowledgments}
We thank N. Armesto for a critical reading of the manuscript.  We
thank the grant Mar\'{\i}a de Maeztu Unit of Excellence of Spain under
project MDM-2016-0692. This work has been funded by Ministerio Ciencia
e Innovaci\'on of Spain under projects FPA2017-83814-P and Xunta de
Galicia (Spain) (Centro singular de investigación de Galicia accreditation
2019-2022) by European Union ERDF, and by  the “María  de Maeztu”  Units  of
Excellence program  MDM-2016-0692. X.F. is supported by grant
ED481B-2019-040 (Xunta de Galicia) and the Fulbright Visiting Scholar
fellowship.

\end{document}